\documentclass{article}
\usepackage{epsf}
\usepackage{times}
\usepackage{pictexwd}
\usepackage{graphicx}
\usepackage{amsmath}
\usepackage{amssymb}
\usepackage{authordate1-4}
\twocolumn
\newcommand{\myspace}{\vspace{0.5cm}}
\newcommand{\myparagraph}[1]{{\myspace \parindent0cm \em #1 \\ \\ \nopagebreak}}
\begin{document}

\title{Electrostatic interactions across a charged lipid bilayer} 

\author{Alexander J.~Wagner and Sylvio May \\ \\
{\small Department of Physics, North Dakota State University, 
Fargo, ND 58105-5566, USA}}
\date{}

\maketitle

\begin{abstract}
{\em 
We present theoretical work in which the degree of electrostatic coupling across a charged lipid bilayer in aqueous solution is analyzed on the basis of nonlinear Poisson-Boltzmann theory. In particular, we consider the electrostatic interaction of a single, large macroion with the two apposed leaflets of an oppositely charged lipid bilayer where the macroion is allowed to optimize its distance to the membrane. Three regimes are identified: a weak and a high macroion charge regime, separated by a regime of close macroion-membrane contact for intermediate charge densities. The corresponding free energies are used to estimate the degree of electrostatic coupling in a lamellar cationic lipid-DNA complex. That is, we calculate to what extent the one-dimensional DNA arrays in a sandwich-like lipoplex interact across the cationic membranes. We find that, in spite of the low dielectric constant inside a lipid membranes, there can be  a significant electrostatic contribution to the experimentally observed cross-bilayer orientational ordering of the DNA arrays. Our approximate analytical model is complemented and supported by numerical calculations of the electrostatic potentials and free energies of the lamellar lipoplex geometry. To this end, we solve the nonlinear Poisson-Boltzmann equation within a unit cell of the lamellar lipoplex using a new lattice Boltzmann method.
}
\end{abstract}

\section*{Introduction}
Lipids self-assemble in aqueous solution into a variety of morphologically 
different structures among which the planar bilayer is the biologically 
most relevant one. Almost all naturally occurring lipid membranes contain 
to some degree acidic lipids which forms the basis for 
electrostatically-driven protein adsorption. In fact, the adsorption of 
macroions onto oppositely charged lipid membranes is an ubiquitous 
phenomenon not only in the living cell but is also common in 
technological and pharmaceutical applications \cite{fevans1}. For example,
the formation of cationic lipid-DNA complexes \cite{PedrosodeLima01}-- 
one of the most promising 
carriers for non-viral gene delivery -- is based on electrostatic 
interactions and subsequent complexation of cationic lipids and 
(the negatively charged) DNA.

What determines the electrostatic properties of a lipid bilayer is not 
only the mole fractions (and valencies) of the charged lipid species
but also the low dielectric constant of the membrane interior \cite{isra1}.
The large drop in the dielectric constant across a lipid bilayer makes it 
energetically unfavorable for large electric fields to form and thus 
dramatically diminishes the electrostatic coupling between the two 
charged monolayers of a lipid bilayer. Similarly, the electrostatic 
interaction between two charged macroions, adsorbed on the opposite 
leaflets of a lipid bilayer, is masked. 

We mention two systems where the electrostatic interaction across a lipid 
bilayer may be significantly enhanced: The first concerns the deposition
near a lipid membrane of an anorganic crystal, calcium pyrophosphate 
dihydrate, leading to crystal-induced membranolysis. A recent Molecular 
Dynamics simulation of crystal deposition on the zwitterionic membrane 
dimyristoylphosphatiadylcholine (DMPC) suggests that membrane rupture
starts in the apposed monolayer and is induced by electrostatic 
interactions across the bilayer \cite{dalal06}.
The second system is the lamellar cationic lipid-DNA complex structure 
\cite{raedler1,lasic2} (also referred to as {\em lipoplex} or $L_\alpha^C$
phase; see below in Fig.~\ref{bx9} for a schematic representation) which 
consists of a lamellar bilayer stack with 
one-dimensional arrays of DNA intercalated. Experimental evidence suggests
that the DNA arrays in different galleries are orientationally ordered
\cite{battersby1}, implying the presence of DNA-DNA interactions across 
the membrane. 
Previous theoretical modeling of this observation was based on the 
assumption that this coupling is mediated through elastic \cite{harries3} or 
electrostatic \cite{schiessel2} interactions. In the latter study, however, 
it was assumed that the membrane is infinitely thin, thus neglecting the 
presence of regions with low dielectric constant. Whether direct 
electrostatic DNA-DNA interaction may lead to the coupling is thus 
still an open question.

The present work analyzes the interaction of a macroion with a charged 
lipid bilayer. Particular focus is put on the role of the low dielectric 
medium inside the membrane and the electrostatic coupling across the bilayer. 
The level of our analysis will be Gouy-Chapman theory which is based on
solving the non-linear Poisson-Boltzmann equation \cite{andelman2}.
Another approximation is that beyond the two apposed 
monolayers of the charged membrane also the membrane-adsorbed 
macroion will be modeled as a very large planar surface. This allows us to 
neglect end-effects and, thus, to solve the {\em one-dimensional} 
Poisson-Boltzmann equation. 
While some aspects of electrostatic interactions across a lipid bilayer 
have previously been analyzed \cite{McQuarrie75,chen4,Genet00,Araghi06} --
mostly on the level of Poisson-Boltzmann theory -- we believe our analysis is 
the first that gives analytical expressions for the free energy 
of the large macroion-bilayer system in all interaction regimes. 

A second objective of the present work is to analyze whether the
electrostatic interactions between the DNA molecules across the cationic 
lipid membranes of a lamellar lipoplex are energetically strong enough 
to induce orientational ordering of the DNA arrays. Based on our free 
energy expressions for macroion-bilayer interactions, we predict that, indeed,
electrostatic interactions are expected to significantly contribute to 
the ordering, at least for conditions of low salt content 
(corresponding to a large Debye length).

Finally, we backup our approximative one-dimensional Poisson-Boltzmann 
model by numerical solutions of the two-dimensional Poisson-Boltzmann 
equation for the lamellar lipoplex geometry shown below 
in Fig.~\ref{bx9} (that is, rigid rods intercalated between a perfectly 
planar membrane stack). Our procedure to numerically solve the 
Poisson-Boltzmann 
equation is a new lattice Boltzmann method which is related to 
a previous application \cite{Horbach01}. The results of our numerical 
calculations are in close agreement with the analytical predictions.

\section*{Electrostatics of a bare membrane}
We start our analysis by considering a charged membrane with no macroions 
present. In this case, it is the two lipid headgroup regions belonging to the
apposed leaflets of the bilayer that may interact electrostatically. 
As outlined in the Introduction, the Poisson-Boltzmann model 
will be used to characterize the electrostatic potential at the headgroups 
and to calculate from that the corresponding free energy.

To facilitate reading of the subsequent parts, we shall first provide 
a summary of the familiar Poisson-Boltzmann model for single charged 
lipid layer. Formally, this case describes an infinitely thick (and thus
electrostatically completely decoupled) bare lipid bilayer.

\myparagraph{Single charged lipid layer}
We consider a single, negatively charged, lipid layer. As is often the case 
in experimental systems, the lipid layer is mixed, containing a monovalently 
charged, and an uncharged (zwitterionic) lipid species. Denoting the mole
fraction of the charged species by $\phi$ and assuming (for the sake of 
simplicity) that both lipid species occupy the same cross-sectional area $a$, 
the surface charge density of the lipid layer is $\sigma=-\phi e/a$ where
$e$ denotes the elementary charge. The minus sign in this equation
accounts for the negative lipid charges.

\begin{figure}[ht] 
\begin{minipage}[t]{6.5cm} 
\includegraphics[width=6.5cm]{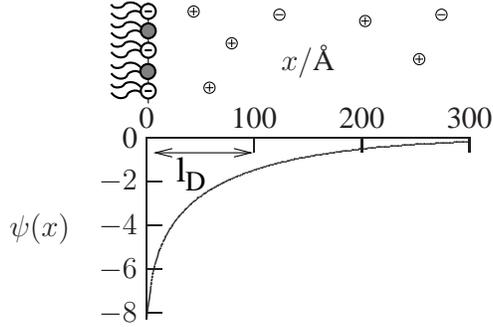}
\end{minipage} 
\caption{\label{hi4} Schematic representation of a binary lipid layer,
consisting of negatively charged and neutral lipids. Some co- and counterions
are depicted. The graph shows the (reduced) electrostatic potential $\psi(x)$
of a single charged lipid layer with composition $\phi=0.5$ and Debye length
$l_D=100\: \mbox{\AA}$. Note that into the calculation 
of $\psi(x)$ 
also enter $l_B=7\:\mbox{\AA}$, and $a=70\:\mbox{\AA}^2$.
The surface potential for this case is $\Psi=-8.28$; see Eq.~\ref{gk2}.} 
\end{figure}

Let us place the lipid layer at the $y,z$-plane of a Cartesian coordinate 
system such that the headgroup charges are located at $x=0$. The aqueous
region extends along the positive $x$-direction ($x>0$) whereas
the lipid tails are directed in the opposite direction ($x<0$) as shown 
in Fig.~\ref{hi4}.
The subject of Poisson-Boltzmann theory is the characterization of the 
small, mobile, ions in the aqueous solution. These ions reflect the presence
of salt with bulk concentration $n_0$. In thermal equilibrium, a diffuse ionic
layer, consisting of co- and counterions, is built that partially
screens the fixed membrane charges. The electrostatic potential 
$\Phi$ can be calculated if the membrane surface charge density, 
$\sigma$, the dielectric constant of water, $\epsilon_W$, and the salt
concentration in the bulk, $n_0$, are known. Note that for the flat, 
large, homogeneous 
lipid layer in Fig.~\ref{hi4}, the electrostatic potential $\Phi=\Phi(x)$ 
depends only on the $x$-direction.
 
In Poisson-Boltzmann theory it is convenient to work, instead of $\Phi$,
$\epsilon_W$, and $n_0$, with related quantities, namely with the 
reduced (dimensionless) electrostatic potential $\psi=e\Phi/k_BT$ (where
$k_B$ is Boltzmann's constant, and $T$ the absolute temperature), with 
the Bjerrum length $l_B=e^2/(4 \pi \epsilon_0 \epsilon_W k_BT)$, and with the 
Debye screening length $l_D=1/\kappa$, defined through $\kappa^2=2 e^2
n_0/(\epsilon_0 \epsilon_W k_BT)$ (where in the latter two definitions
$\epsilon_0$ is the permittivity of free space). 
In terms of these quantities, Poisson-Boltzmann theory leads to the 
Poisson-Boltzmann equation $\psi''(x)=\kappa^2 \sinh \psi(x)$  that the
reduced potential $\psi(x)$ has to fulfill subject to the two boundary
conditions $\psi(x \to \infty)=0$ and $\psi'(0)=2 \kappa p$
where we have used the short notation $p=p_0 \phi$ with the constant 
$p_0=2 \pi l_B l_D/a$.
(Note that a prime denotes the derivative; i.~e., $\psi'(x)=d
\psi(x)/dx$ etc.)  Integrating the Poisson-Boltzmann equation once 
leads with $\psi(x \to \infty)=0$ to 
\begin{equation} \label{dd2}
\psi'(x)=-2 \kappa \: \sinh \left(\frac{\psi(x)}{2}\right)
\end{equation}
After a second integration we obtain the 
final solution of the Poisson-Boltzmann equation for a single monolayer
\begin{equation} \label{fr43}
\psi(x)=-2 \ln \left\{ 1+\frac{2}{e^{\kappa x} 
\coth\left[\mbox{arsinh}(p)/2\right]-1}\right\}
\end{equation}
which is plotted for one characteristic case in Fig.~\ref{hi4}.
We note that from Eq.~\ref{fr43} one may obtain the surface potential 
$\Psi=\psi(x=0)$. The result is
\begin{equation} \label{gk2}
\Psi=-2 \: \mbox{arsinh} (p_0 \phi)
\end{equation}
For later use we also note that in terms of the surface potential 
$\Psi$ the potential $\psi(x)$ is given by
\begin{equation} \label{fr4}
\psi(x)=-2 \ln \left( \frac{e^{\kappa x}-\tanh (\Psi/4)}{e^{\kappa x}+
\tanh (\Psi/4)} \right)
\end{equation}
Poisson-Boltzmann theory also predicts that the relation between the 
electrostatic potential and surface charge density can be used to 
calculate the free energy per lipid through a {\em charging process} 
\cite{fevans1}
\begin{equation} \label{hr6}
f_{ml}=a \: \int \limits_0^\sigma \bar{\Phi}(\tilde{\sigma}) \: d\tilde{\sigma}
=-k_BT \: \int \limits_0^\phi \Psi(\tilde{\phi}) \: d\tilde{\phi}
\end{equation}
where $\bar{\Phi}=\Phi(x=0)$ is the electrostatic 
surface potential. The integration can be carried out because 
the relation $\Psi(\phi)$ is given in Eq.~\ref{gk2}; the result is
\begin{equation} \label{hr8}
f_{ml}(\phi)=2 \phi \left[\frac{1-q}{p}+\ln (p+q)\right]
\end{equation}
with $q=\sqrt{p^2+1}$ (recall $p=\phi p_0$).
Note that here and in the following we shall express all energies in 
units of $k_BT$.
Eq.~\ref{hr8} requires $\phi \geq 0$. Yet, the convention 
$f_{ml}(-\phi)=f_{ml}(\phi)$ ensures applicability of Eq.~\ref{hr8}
also for positively charged surfaces where $\phi<0$.
Note finally that Eq.~\ref{hr6} implies $f_{ml}'(\phi)=-\Psi(\phi)$.

\myparagraph{Charged lipid bilayer}
A lipid bilayer consists of two apposed lipid monolayers that form a thin film
of low dielectric constant. The electrostatic properties on one side of the
bilayer may, in principle, depend on those of the other side. This coupling is 
the subject of the present paragraph. 

\begin{figure}[ht]
\begin{center}
\includegraphics[width=8.5cm]{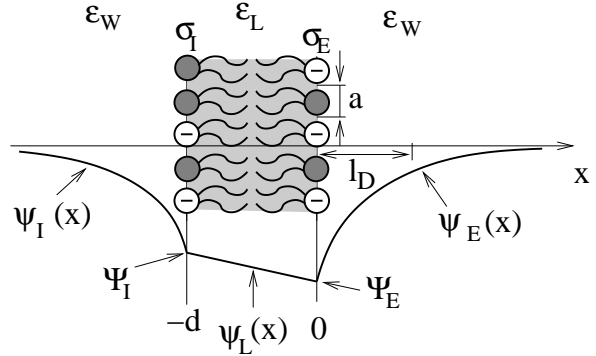} 
\end{center} 
\caption{\label{cc4} Schematic illustration the lipid bilayer, located within
the region $-d < x \leq 0$. The reduced (dimensionless) electrostatic 
potentials
are $\psi_I(x)$ and $\psi_E(x)$ in the aqueous regions, and $\psi_L(x)$
between the charged monolayers, each of surface charge density, 
$\sigma_I=-e \phi_I/a$ and $\sigma_E=-e \phi_E/a$. The reduced
surface potentials are denoted by $\Psi_I$ and $\Psi_E$.} 
\end{figure}

In the following, we shall refer
to the two apposed monolayers of a lipid bilayer as {\em external} and 
{\em internal}) and denote their compositions by $\phi_E$ and 
$\phi_I$, respectively. Fig.~\ref{cc4} illustrates 
\footnote{fig 2 here}
a number of additional 
structural properties: The lipid bilayer has hydrophobic thickness $d$ with  
the charged headgroups of its two monolayers being located at $x=0$ and 
$x=-d$, and the dielectric constant in the inner hydrophobic region 
($-d \leq x \leq 0$) is $\epsilon_L$.
There are three different regions in which we denote the reduced 
electrostatic potential, $\psi(x)$,
by $\psi_I(x)$ (for $-\infty < x \leq -d$), by  $\psi_L(x)$ 
(for $-d < x \leq 0$), and by $\psi_E(x)$ (for $0 < x \leq \infty$).
The reduced potential then fulfills the Poisson-Boltzmann equation
in the aqueous regions adjacent to each monolayer and the Laplace equation 
inside the membrane
\begin{eqnarray} \label{ui8}
\psi_I''(x)&=&\kappa^2 \sinh \psi_I \hspace{1cm} x \leq -d \nonumber\\
\psi_L''(x)&=&0 \hspace{1.6cm} -d < x < 0 \\
\psi_E''(x)&=&\kappa^2 \sinh \psi_E \hspace{1cm} 0 \leq x \nonumber
\end{eqnarray}
It is convenient to define a coupling parameter  
\cite{Winterhalter92} through
\begin{equation}
H=\frac{\epsilon_L l_D}{\epsilon_W d} 
\end{equation}
With that, the boundary conditions at the two interfacial regions read
\begin{eqnarray} \label{wd4}
\psi_E'(0)-H d \kappa \psi_L'(0)&=&2 \kappa p_0 \phi_E\nonumber\\
-\psi_I'(-d)+H d \kappa \psi_L'(-d)&=&2 \kappa p_0 \phi_I
\end{eqnarray}
Note that, as we expect, for $H=0$ the two conditions reduce to the
corresponding ones for two single, electrostatically decoupled, monolayers. 
The potential between the two charged monolayers fulfills a
one-dimensional Laplace equation and is thus given by the linear function, 
\begin{equation}
\psi_L(x)=\Psi_I+\left(1+\frac{x}{d}\right) (\Psi_E-\Psi_I)
\end{equation}
where $\Psi_I=\psi_I(-d)=\psi_L(-d)$ and
$\Psi_E=\psi_L(0)=\psi_E(0)$ are the surface potentials at the 
internal and external monolayer, respectively. We thus have 
$\psi_L'=\triangle \Psi/d$ where 
\begin{equation}
\triangle \Psi=\Psi_E-\Psi_I 
\end{equation}
is the difference between the reduced electrostatic potentials  
across the bilayer. 
From the first integration of the Poisson-Boltzmann 
equation (see Eq.~\ref{dd2})
and the boundary conditions (see Eqs.~\ref{wd4}) we find the surface potentials
\begin{equation} \label{zef}
\Psi_E=-2 \mbox{arsinh} (p_0 \bar{\phi}_E) \hspace{1cm}
\Psi_I=-2 \mbox{arsinh} (p_0 \bar{\phi}_I)
\end{equation}
where we have introduced the effective compositions
\begin{equation} \label{vt5}
\bar{\phi}_E=\phi_E+\frac{H \triangle \Psi}{2 p_0} \hspace{1cm}
\bar{\phi}_I=\phi_I-\frac{H \triangle \Psi}{2 p_0} 
\end{equation}
We note that the surface potentials, $\Psi_E=\Psi_E(\phi_E,\phi_I)$
and $\Psi_E=\Psi_E(\phi_E,\phi_I)$, depend on both monolayer
compositions, $\phi_E$ and $\phi_I$, because of the electrostatic coupling
between the charged monolayers.
Combination of Eq.~\ref{zef} and Eq.~\ref{vt5} leads to 
$\triangle \Psi=f_{ml}'(\bar{\phi}_I)-f_{ml}'(\bar{\phi}_E)$ or, equivalently,
\begin{equation} \label{tg1}
\frac{\triangle \Psi}{2}=
\mbox{arsinh} \left(p_0 \phi_I-\frac{H}{2} \triangle \Psi \right)-
\mbox{arsinh} \left(p_0 \phi_E+\frac{H}{2} \triangle \Psi\right) 
\end{equation}
which represents a transcendental equation for the potential difference 
$\triangle \Psi$ across the membrane \cite{baciu04}. Of course, for a symmetric bilayer,
where $\phi_E=\phi_I$, the solution of Eq.~\ref{tg1} is $\triangle \Psi=0$.

The electrostatic free energy, $f_{bl}$ (measured per lipid pair)
is given by a charging process analogous to Eq.~\ref{hr6}
\begin{equation} \label{he5}
f_{bl}=-\int \limits_0^{\phi_I} \Psi_I(\tilde{\phi}_I,0) d\tilde{\phi}_I
-\int \limits_0^{\phi_E} \Psi_E(\phi_I,\tilde{\phi}_E) d\tilde{\phi}_E
\end{equation}
where here we first charge the internal monolayer, and then the external one
(any other path of charging would lead to the same result). We obtain
\begin{equation} \label{jj8}
\fbox{$ \displaystyle
f_{bl}=f_{ml}(\bar{\phi}_E)+f_{ml}(\bar{\phi}_I)+
\frac{H}{4 p_0} \triangle \Psi^2
$}
\end{equation}
The first two terms represent the charging energies for two single 
monolayers of effective compositions $\bar{\phi}_E$ and $\bar{\phi}_I$. 
The last term, 
$H \triangle \Psi^2/4 p_0=\epsilon_L a \triangle \bar{\Phi}^2/2 d$ 
is the contribution of the capacitor between the charged monolayer 
interfaces where $\triangle \bar{\Phi}$ is the potential difference between 
the external and internal monolayer.

For $H=0$ there is no electrostatic coupling between the charged monolayers.
In this case, $f_{bl}=f_{ml}(\phi_E)+f_{ml}(\phi_I)$ is given by the sum of 
the individual free energies of two isolated monolayers.

\section*{Macroion-bilayer interaction}
After heaving dealt with a bare lipid bilayer, we are now in the position to 
add a positively charged macroion to the external side of the bilayer.
To be specific, we place the charges of the macroion at position $h$
along the $x$-axis as shown in Fig.~\ref{cc1} below.
It is convenient to express the charge density, $\sigma_P=\phi_P e/a$, 
of the macroion in terms of the compositional variable $\phi_P$ (with $0 \leq
\phi_P \leq 1$). For $\phi_P=\phi_E$ the charge densities on the macroion and on
the external monolayer match each other; that is, they would be opposite in
sign and equal in magnitude.

As for the isolated bilayer, the potentials, $\psi_I(x)$, $\psi_L(x)$, and
$\psi_E(x)$, are determined by Eqs.~\ref{ui8}. The only difference is that 
the last one of Eqs.~\ref{ui8} applies only within the region, 
$0 \leq x \leq h$, between the external monolayer and the macroion. 
The boundary condition at the macroion is $\psi_E'(h)=2 \kappa p_0 \phi_P$.
The other two boundary conditions (see Eq.~\ref{wd4}) remain unchanged.
The surface potentials $\Psi_I=\psi_I(-d)=\psi_L(-d)$, 
$\Psi_E=\psi_L(0)=\psi_E(0)$, and $\Psi_P=\psi_E(h)$ depend
on the three compositions, $\phi_I$, $\phi_E$, and $\phi_P$. We thus write
$\Psi_I=\Psi_I(\phi_I,\phi_E,\phi_P)$,
$\Psi_E=\Psi_E(\phi_I,\phi_E,\phi_P)$, and 
$\Psi_P=\Psi_P(\phi_I,\phi_E,\phi_P)$.
Fig.~\ref{cc1} schematically illustrates the macroion-bilayer system.
\begin{figure}[ht]
\begin{center}
\includegraphics[width=7.5cm]{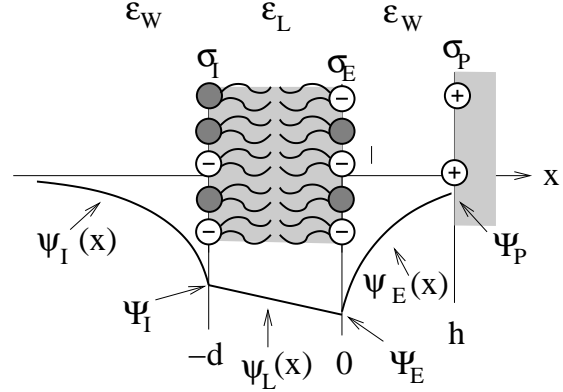} 
\end{center} 
\caption{\label{cc1} Schematic illustration the lipid bilayer, 
interacting with a macroion. The difference to Fig.~\ref{cc4} is the presence
of the positively charged macroion of surface charge density $\sigma_P=e
\phi_P/a$ at position $x=h$. The surface potential at
the macroion is denoted by $\Psi_P$. (Depicted is the case of a weakly 
charged macroion.)} 
\end{figure}
The free energy of the macroion-membrane system, measured per lipid in
one monolayer, is again given by the familiar charging process
analogous to Eq.~\ref{he5}
\begin{eqnarray} \label{lz2}
f&=&-\int \limits_0^{\phi_I} \Psi_I(\tilde{\phi}_I,0,0) d\tilde{\phi}_I
-\int \limits_0^{\phi_E} \Psi_E(\phi_I,\tilde{\phi}_E,0) d\tilde{\phi}_E
\nonumber\\
&+&\int \limits_0^{\phi_P} \Psi_P(\phi_I,\phi_E,\tilde{\phi}_P) d\tilde{\phi}_P
\end{eqnarray}
It corresponds to charging first the internal monolayer (first term in 
Eq.~\ref{lz2}), then the external monolayer (second term in Eq.~\ref{lz2}), 
and finally the macroion (last term in Eq.~\ref{lz2}).
We repeat that the final result for $f$ does of course not depend on the 
order in which the surfaces are charged.

In the following, the distance $h$ between the macroion and the
bilayer is chosen such that the free energy $f$ is minimal. 
In this case, the potential $\psi_E(x)$ has the same functional form as for an
isolated charged monolayer \cite{gingell1}: 
for small $\phi_P$ the potential is that of the
isolated lipid bilayer, and for large $\phi_P$ the potential is determined
solely by the macroion charges. The small and large $\phi_P$ cases are 
separated by another, intermediate, regime where the macroion contacts the 
bilayer (implying $h=0$). In the following, we shall consider 
the three possible scenarios separately.

\myparagraph{Weakly charged macroion}
If $\phi_P$ is sufficiently low the potential $\psi(x)$ is equivalent to that
of an isolated bilayer within the region $-\infty \leq x \leq h$. The surface
potentials at the membrane are thus, again, given by Eq.~\ref{zef},
and the value of the potential at the macroion, 
\begin{equation}
\Psi_P=-2 \mbox{arsinh} (p_0 \phi_P) 
\end{equation}
is negative. Recalling the functional form of 
Eq.~\ref{fr4} and $\psi_E(h)=\Psi_P$, it follows that
\begin{equation} \label{nk2}
e^{\kappa h}=\frac{\tanh \left(\Psi_E/4\right)}
{\tanh \left(\Psi_P/4\right)}=\left(\frac{\phi_P}{\bar{\phi}_E}\right) \:
\left(\frac{\sqrt{1+(p_0 \bar{\phi}_E)^2}-1}{\sqrt{1+(p_0 \phi_P)^2}-1}\right)
\end{equation}
which allows us to calculate the equilibrium distance, $h$, between the 
macroion and the external monolayer. With growing $\phi_P$ the distance $h$ 
decreases until close contact is established for a particular 
$\phi_P$ which fulfills the relation $\phi_P=\bar{\phi}_E$. The condition  
\begin{equation} \label{gr5}
\phi_P < \phi_E+\frac{H}{2 p_0} \triangle \Psi
\end{equation}
thus defines the {\em weakly} charged macroion regime. Recall that the 
potential difference, $\triangle \Psi$, fulfills Eq.~\ref{tg1}. 
The presence of the oppositely charged macroion reduces the free energy of the
isolated bilayer by the charging free energy of the isolated macroion
\begin{equation} \label{lf6}
\fbox{$ \displaystyle
f=f_{ml}(\bar{\phi}_E)+f_{ml}(\bar{\phi}_I)+ 
\frac{H}{4 p_0} \triangle \Psi^2-f_{ml}(\phi_P)
$}
\end{equation}

\myparagraph{Highly charged macroion}
For sufficiently high $\phi_P$, the potential between the macroion and 
the external monolayer ($0 \leq x \leq h$) is equal to that of the isolated 
macroion, implying
\begin{equation} \label{lu8}
\psi_E(x)=-2 \ln \left( \frac{e^{-\kappa (x-h)}-\tanh (\Psi_P/4)}
{e^{-\kappa (x-h)}+\tanh (\Psi_P/4)} \right)
\end{equation}
Hence, both the surface potentials at the macroion and at the external
monolayer are positive, 
\begin{equation} \label{gj6}
\Psi_E=2 \mbox{arsinh} (p_0 \bar{\phi}_E) \hspace{1cm}
\Psi_P=2 \: \mbox{arsinh} (p_0 \phi_P)
\end{equation}
whereas the surface potential of the internal monolayer
\begin{equation} \label{dj6}
\Psi_I=-2 \mbox{arsinh} (p_0 \bar{\phi}_I)
\end{equation}
can be either negative ($\bar{\phi}_I>0$) or positive ($\bar{\phi}_I<0$), 
depending on the sign of $\bar{\phi}_I$. Inserting the definitions for 
$\bar{\phi}_I$ and $\bar{\phi}_E$ (see Eqs.~\ref{vt5}) we find a
transcendental equation for the potential difference $\triangle \Psi$
across the bilayer
\begin{equation} \label{tg2}
\frac{\triangle \Psi}{2}=
\mbox{arsinh} \left(p_0 \phi_I-\frac{H}{2} \triangle \Psi \right)+
\mbox{arsinh} \left(p_0 \phi_E+\frac{H}{2} \triangle \Psi\right) 
\end{equation}
Because of $\Psi_E=\psi_E(0)$ according to Eq.~\ref{lu8}, and with 
Eq.~\ref{gj6}, we can determine the equilibrium distance $h$ through
\begin{equation} \label{ji5}
e^{\kappa h}=\frac{\tanh \left(\Psi_P/4\right)}
{\tanh \left(\Psi_E/4\right)}=\left(\frac{\bar{\phi}_E}{\phi_P}\right) \:
\left(\frac{\sqrt{1+(p_0 \phi_P)^2}-1}{\sqrt{1+(p_0 \bar{\phi}_E)^2}-1}\right)
\end{equation}
With decreasing $\phi_P$ the distance $h$ decreases too until
close contact is established for a particular $\phi_P$ which
fulfills the relation $\phi_P=\bar{\phi}_E$.
Hence, the {\em highly} charged macroion regime is defined by the condition
\begin{equation} \label{nd4}
\phi_P > \phi_E+\frac{H}{2 p_0} \triangle \Psi
\end{equation}
where, the potential difference, $\triangle \Psi$, results from
the solution of Eq.~\ref{tg2}. Note that the surface potential at the internal
monolayer, $\Psi_I$, is positive if
$H \triangle \Psi > 2 p_0 \phi_I$, and combination with Eq.~\ref{nd4}
reveals that $\phi_P>\phi_E+\phi_I$ is a necessary condition to reverse the
sign of both surface potentials, $\Psi_I$ and $\Psi_E$.

The presence of the macroion leads to the free energy
\begin{equation} \label{xl6}
\fbox{$ \displaystyle
f=-f_{ml}(\bar{\phi}_E)+f_{ml}(\bar{\phi}_I)+
\frac{H}{4 p_0} \triangle \Psi^2+f_{ml}(\phi_P)
$}
\end{equation}

\myparagraph{Intermediate case: Close contact}
In between the weakly and highly charged macroion regimes (defined by 
Eq.~\ref{gr5} and Eq.~\ref{nd4}) the macroion and the external monolayer 
are at close contact.
Here, the surface potentials at the lipid bilayer are determined by the
equations 
\begin{eqnarray} \label{vn5}
\Psi_I&=&-2 \: \mbox{arsinh} [p_0 (\phi_I+\phi_E-\phi_P)]\\
\Psi_E&=&-2 \: \mbox{arsinh} [p_0 (\phi_I+\phi_E-\phi_P)]-\frac{2 p_0}{H}
(\phi_E-\phi_I)\nonumber
\end{eqnarray}
Note that because of $h=0$ it is $\Psi_E=\Psi_P$.
The free energy can be calculated by the usual charging process
\begin{equation}
f= -\int \limits_0^{\phi_I} \Psi_I(\tilde{\phi}_I,\phi_E-\phi_P=0) \:
d\tilde{\phi}_I 
-\int \limits_0^{\phi_E-\phi_P} \Psi_E(\phi_I,\tilde{\phi}) \:
d\tilde{\phi}
\end{equation} 
which corresponds to first charging the internal monolayer and then the 
together the external layer and the macroion. We obtain the result
\begin{equation} \label{db6}
\fbox{$ \displaystyle
f=f_{ml}(\phi_I+\phi_E-\phi_P)+\frac{p_0}{H} {(\phi_E-\phi_P)}^2
$}
\end{equation}
which is the sum of the electrostatic free energy of a single isolated
monolayer 
of composition $|\phi_I+\phi_E-\phi_P|$ and the capacitor energy 
$H \triangle \Psi^2/4 p_0$.
Note the particular case $\phi_P=\phi_E+\phi_I$ for which $f=p_0 \phi_I^2/H$.

\section*{Two limiting cases}
The free energies $f$ in Eqs.~\ref{lf6}, \ref{xl6} and \ref{db6} cover all
possible interaction regimes of a macroion with an oppositely charged 
lipid bilayer.
The actual calculation of $f$, however, requires to solve an additional
transcendental equation for $\triangle \Psi$. There are two limiting cases
for which $f$ can be  expressed directly in terms of the membrane and macroion
charge densities. One is the Debye-H\"uckel limit which is based on
linearizing the Poisson-Boltzmann equation. The other one corresponds to a 
very thin (transparent) lipid bilayer for which the limit 
$d \to 0$ may be considered. These two cases are discussed 
in the following.

\myparagraph{Debye-H\"uckel limit}
For sufficiently small charge densities the Poisson-Boltzmann 
equation can be linearized.
Formally, the Debye-H\"uckel case may be viewed as the limit corresponding 
to $l_D \to 0$ (or, equivalently, $p_0 \to 0)$.
In this case, all three regimes of the interacting macroion-bilayer system
can be calculated explicitly.

In the {\em weakly charged macroion} regime linearization of Eq.~\ref{tg1} 
gives rise to
\begin{equation} \label{fa7}
\triangle \Psi=-2 p_0 \: \frac{\phi_E-\phi_I}{1+2 H}
\end{equation}
We can use this result to calculate the condition for the 
system to be in the weakly charged macroion regime (see Eq.~\ref{gr5})
\begin{equation}
\phi_P < \phi_E-\frac{H}{1+2 H} \: (\phi_E-\phi_I)
\end{equation}
as well as the relation for the distance, $h$, between the macroion and 
the external monolayer (see Eq.~\ref{nk2})
\begin{equation}
e^{\kappa h}=\frac{\phi_E+H (\phi_E+\phi_I)}{\phi_P (1+2 H)}
\end{equation}
and the free energy (see Eq.~\ref{lf6})
\begin{equation} \label{ke5}
\frac{f}{p_0}=\phi_I^2+\phi_E^2-\phi_P^2-\frac{H}{1+2 H} (\phi_E-\phi_I)^2
\end{equation}

Similarly in the {\em highly charged macroion} regime, linearization of
Eq.~\ref{tg2} results in
\begin{equation}
\triangle \Psi=2 p_0 (\phi_E+\phi_I)
\end{equation}
This result is used to calculate the condition for the system to reside in the
highly charged macroion regime (see Eq.~\ref{nd4})
\begin{equation}
\phi_P > \phi_E+H (\phi_E+\phi_I)
\end{equation}
as well as the relation for the distance, $h$, between the 
macroion and the external monolayer (see Eq.~\ref{ji5})
\begin{equation}
e^{\kappa h}=\frac{\phi_P}{\phi_E+H (\phi_E+\phi_I)}
\end{equation}
and the free energy (see Eq.~\ref{xl6})
\begin{equation} \label{jr9}
\frac{f}{p_0}=\phi_I^2-\phi_E^2+\phi_P^2-H (\phi_E+\phi_I)^2
\end{equation}

Finally, for the {\em intermediate case} of close contact, Eq.~\ref{db6}
already provides an explicit expression which in the  Debye-H\"uckel limit
reads 
\begin{equation} \label{jr6}
\frac{f}{p_0}=(\phi_I+\phi_E-\phi_P)^2+\frac{1}{H} \: (\phi_E-\phi_P)^2
\end{equation}
Of course, at the transition from one regime into the other the free energy
is continuous (but not necessarily smooth). That is, 
for $\phi_P=\phi_E-(\phi_E-\phi_I) H/(1+2 H)$, the free
energy $f$ in Eq.~\ref{ke5} is equal to that in Eq.~\ref{jr6}. Similarly,
for $\phi_P=\phi_E+H (\phi_E+\phi_I)$, the free
energy $f$ in Eq.~\ref{jr9} is equal to that in Eq.~\ref{jr6}.

\myparagraph{Transparent bilayer}
For very thin membranes ($d \to 0$) or in the limit of large Debye
length ($l_D \to \infty$) it is $H \to \infty$, 
and the membrane appears transparent to the electric field. 

Consider first the low macroion charge limit. For $H \to \infty$
we obtain from  Eq.~\ref{tg1} the potential difference
\begin{equation}
\triangle \Psi=-\frac{p_0}{H} \: (\phi_E-\phi_I) \to 0
\end{equation}
The effective compositions, defined in Eq.~\ref{vt5}, are then simply
\begin{equation}
\bar{\phi}_E=\bar{\phi}_I=\frac{1}{2} \: (\phi_E+\phi_I)
\end{equation}
We also see that the condition for residing in the weakly charged macroion
regime becomes $\phi_P<\bar{\phi}$ where we have defined the average 
composition of
the lipid bilayer $\bar{\phi}=(\phi_I+\phi_E)/2$. The distance between the
macroion and the external monolayer is again given by Eq.~\ref{nk2} with
$\bar{\phi}_E=\bar{\phi}$.
Finally, the free energy (see Eq.~\ref{lf6}) becomes
\begin{equation}
f=2 f_{ml}(\bar{\phi})-f_{ml}(\phi_P)
\end{equation}
For $\phi_P>\bar{\phi}$ there is close contact between the external monolayer
and the macroion. A third, high macroion charge regime, does
not exist for a transparent membrane. In the close contact regime, 
Eq.~\ref{db6} predicts $f=f_{ml}(\phi_I+\phi_E-\phi_P)$.

\section*{Electrostatic coupling between the DNA monolayers
in lamellar lipoplexes}
\begin{figure}[ht]
\begin{center}
\begin{minipage}[b]{6.0cm} 
\includegraphics[width=6.0cm]{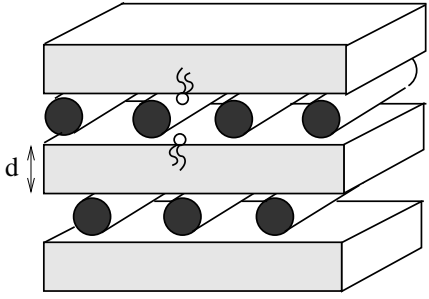} 
\end{minipage}
\caption{\label{bx9} Schematic illustration of the lamellar lipoplex 
structure. The (double stranded) DNA molecules, depicted as long rods,
form one-dimensional arrays between each pair of cationic bilayers.
Arrays in consecutive membrane galleries are orientationally aligned
and shifted ''out of phase''. Some lipids are schematically shown; 
$d$ denotes the membrane thickness.} 
\end{center} 
\end{figure}

In this section we shall apply our results for the electrostatic coupling 
to a specific system, the lamellar lipoplex. As mentioned in the 
Introduction, there is experimental evidence for the orientational 
alignment of the DNA arrays across the cationic bilayers \cite{battersby1}.
Such an alignment implies some kind of energetic coupling. Our results 
from the previous sections enable us to calculate the 
electrostatic contribution to this coupling. 
Fig.~\ref{bx9} depicts\footnote{fig 4 here}
the structure of the lamellar 
lipoplex with its one-dimensional arrays of DNA monolayers being
intercalated between a stack of cationic bilayers.

\begin{figure}[ht]
\begin{center}
\begin{minipage}[b]{8.0cm}
\includegraphics[width=8.0cm]{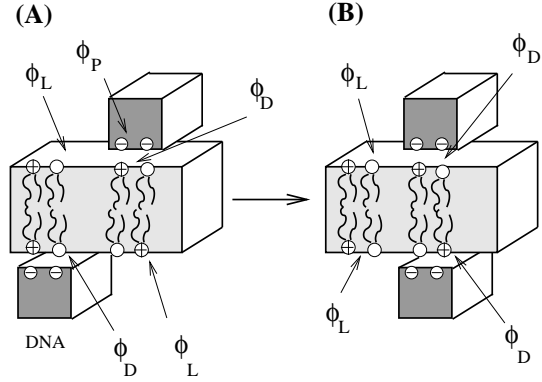} 
\end{minipage}
\caption{\label{bx8} Our structural model to calculate the degree of 
electrostatic coupling of two DNA rods (depicted with rectangular 
cross-section) across a cationic bilayer:
in state (A) the DNA rods are ''out of phase'' so that potential gradients 
along the membrane normal direction can form. In state (B) the 
DNA rods are ''in phase'', implying transmembrane symmetry and thus
vanishing gradients. The difference in free energy of states (A) and (B) 
is a measure for the degree of electrostatic coupling. Note that, 
in general, the mixed cationic membrane may optimize its composition (mole 
fraction of charged lipids); $\phi_D$ and $\phi_L$ denote the compositions
at the DNA adsorption region and at the bare membrane, respectively.
} 
\end{center} 
\end{figure}

Our objective is to use the formalism developed in the previous sections 
to (roughly) estimate the interaction strength of the DNA molecules through 
a charged lipid bilayer. To do so, it is convenient to consider only a single,
perfectly planar, cationic lipid bilayer with two oppositely charged DNA
arrays attached to each of its apposed monolayers. Assuming all DNA strands 
to be parallel we shift one array with respect to the other one. 
The corresponding free energy is expected to adopt its minimum and maximum 
for the two DNA arrays being ''out of phase'' and ''in phase'', respectively 
(see states A and B in  Fig.~\ref{bx8}).

At this point it should be noted that the recent theoretical prediction of 
a so-called {\em sliding columnar phase} is based not only on an energetic
penalty of laterally shifting the two DNA arrays 
(as shown in Fig.~\ref{bx8}) but also on a second energy contribution 
associated with angular changes of the DNA arrays with respect to each 
other \cite{ohern1,golubovic1}. Clearly, our present simplified model allows
to estimate the former but not the latter contribution.

We calculate the corresponding free energy difference, $\triangle F$, between 
states A and B per persistence length, $\xi=500 \: \mbox{\AA}$, of double
stranded DNA.
Let us denote by $N$ the number of lipids that interact
electrostatically with the DNA. There is some arbitrariness in choosing $N$
because the planar macroion geometry (see Fig.~\ref{bx8})
is not a very realistic model for a DNA
molecule. Still a reasonable estimate is $N=2 \xi R_D/a \approx 150$ where
$R_D=10 \: \mbox{\AA}$ is the radius of the DNA rod (typically, 
$a\approx 70 \: \mbox{\AA}^2$ is the cross-sectional area per lipid).
The additional factor of two in $N=2 \xi R_D/a$ results from the two monolayers
that interact with a single DNA molecule.

The main difference of states A and B in Fig.~\ref{bx8} is the symmetry
of state B with respect to the bilayer's midplane. That is, for state B the
potential difference across the bilayer $\triangle \Psi$ vanishes 
(corresponding to $H=0$) and we can
write for the difference in free energy between state A and B
\begin{equation} \label{lo3}
\triangle F=F_A-F_B=N [f(\phi,H)-f(\phi,H=0)]
\end{equation}
where $f$ is the free energy per lipid in one monolayer, and is given 
by Eq.~\ref{lf6} for low macroion charge, by
Eq.~\ref{db6} for close contact, and by Eq.~\ref{xl6} for high macroion charge.

To specify the compositional variable $\phi$ in Eq.~\ref{lo3}
we note the ability of the lipid bilayer to adjust its local lipid
composition. That is, close to the DNA adsorption site the local
lipid composition, which we shall denote by $\phi_D$, may differ from that at
the DNA-free bilayer region (denoted by $\phi_L$ in the following).
The ability for local demixing can (approximatively) be included in 
our calculation of $\triangle F$.
In order to derive an explicit relation we shall assume small electrostatic
coupling $H \ll 1$ for which we can expand $\triangle F$ up to first order in
$H$. The result is 
\begin{equation} \label{ko5}
\triangle F=-\frac{N H}{4 p_0} \: [\triangle \Psi(\phi_D,\phi_L)]^2=-\frac{N H}{4 p_0} \: \left[f_{ml}'(\phi_D)\mp f_{ml}'(\phi_L) 
\right]^2 
\end{equation}
where the minus sign refers to the low and the plus sign to the high macroion
charge regime (in the small $H$ limit the small macroion charge regime is 
$\phi_P<\phi_D$ and the high macroion charge regime is $\phi_P>\phi_D$;
the close contact regime reduces to $\phi_P=\phi_D$).

Let us analyze the consequences of Eq.~\ref{ko5}. For frozen lipid composition,
where $\phi_D=\phi_L=\phi$, we find $\triangle F=0$ in the low macroion charge
regime. This is expected because for low $\phi_P$ the macroions do not affect
the membrane potentials. Yet, the membrane is symmetric, implying 
$\triangle \Psi=0$, and the electrostatic coupling across the bilayer
is irrelevant.

\begin{figure}[ht] 
\begin{center} 
\begin{minipage}[t]{7cm} 
\includegraphics[width=6cm]{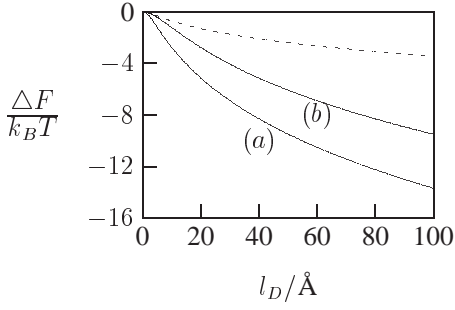}
\end{minipage}
\caption{\label{xa2} $\triangle F$ according to Eq.~\ref{ko6} 
for $l_B=7\:\mbox{\AA}$, $a=70\:\mbox{\AA}^2$,
$N=80$, $\epsilon_W/\epsilon_L=40$, and $d=30\:\mbox{\AA}$ as a function of the
Debye length $l_D$. The two curves correspond to $\phi=0.5$ (a) and 
$\phi=0.25$, (b). The broken line shows $\triangle F$ according to
Eq.~\ref{jt6} for $\phi_P=0.5$.}
\end{center} 
\end{figure}

Consider now the high macroion charge
regime for frozen lipid composition ($\phi_D=\phi_L=\phi$). Here 
Eq.~\ref{ko5} becomes
\begin{equation} \label{ko6}
\triangle F=-4 N \frac{H}{p_0} \: \mbox{arsinh}^2 (p_0 \phi)
\end{equation}
In Fig.~\ref{xa2} we plot $\triangle F$ according to Eq.~\ref{ko6}
(with $H=\epsilon_L l_D/\epsilon_W d$ and $p_0=2 \pi l_D l_B/a$)
for $l_B=7\:\mbox{\AA}$, $a=70\:\mbox{\AA}^2$, $N=150$,
$\epsilon_W/\epsilon_L=40$, and $d=30\:\mbox{\AA}$ as a function of the
Debye length $l_D$. Curve (a) is derived for $\phi=0.5$, curve (b) for
$\phi=0.25$. In both cases $\triangle F$ can become considerably larger than
$k_BT$ under low salt conditions.

Let us also estimate the influence of local lipid demixing.
To this end we consider a weakly charged membrane. The presence of the DNA's 
induces accumulation of charged lipids near the DNA. From an electrostatic
point of view, local charge matching ($\phi_D \approx \phi_P$) would be the
optimal situation. The matching $\phi_D=\phi_P$ would be opposed by the
demixing free energy of the lipids but we may still consider the limiting
situation that all charged lipids are driven towards the DNA adsorption site,
implying $\phi_D=\phi_P$ and $\phi_L=0$. Eq.~\ref{ko5} then reads
\begin{equation} \label{jt6}
\triangle F=-N \frac{H}{p_0} \: \mbox{arsinh}^2 (p_0 \phi_P)
\end{equation}
Comparison of Eq.~\ref{ko6} with Eq.~\ref{jt6} shows that the lateral mobility
of the lipids tends to reduce the electrostatic coupling of the DNA arrays 
through the cationic bilayer.
Sufficiently strong electrostatic coupling is generally found only if the Debye
length is considerably larger than that under physiological conditions where
$l_D \gg 10 \: \mbox{\AA}$.


\begin{figure}
\begin{center}
\includegraphics[width=8cm]{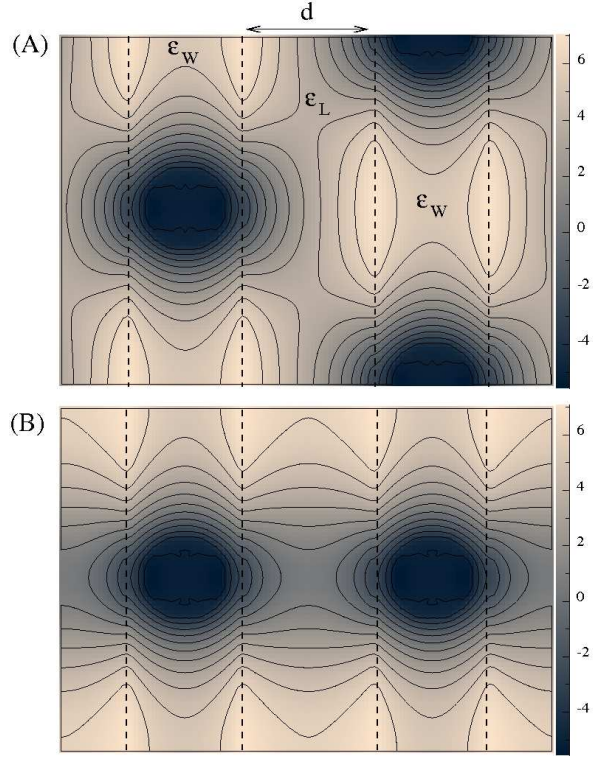}
\end{center}
\caption{\label{FigSIM} Simulation results for the potential $\Psi({\bf r})$
in a lamellar lipoplex. The two diagrams correspond to the ''out of phase'' (A)
and ''in phase'' (B) cases (see also Fig.~\ref{bx8}).   
Contour lines are drawn at integer values of $\Psi$. The simulation is 
carried out for a 112 \AA x 160 \AA system. The membrane width 
is $d=30 \mbox{\AA}$, the DNA radius is $R_D=10 \mbox{\AA}$, and 
the distance between the DNA and the membrane is $3 \mbox{\AA}$. 
We use $l_D=100 \mbox{\AA}$, $l_B=7 \mbox{\AA}$ and 
$\epsilon_L/ \epsilon_W= 1/40$. The 
membrane is homogeneously charged  with composition $\phi=0.25$ of 
monovalent cationic lipids. The charge density on the DNA is twice as large 
as that on the membrane, amounting to a linear charge 
density of $-e/1.7 \mbox{\AA}$ along each DNA rod.
The location of the membrane charges is indicated by the vertical 
dashed lines.}
\end{figure}

\section*{Solving the Poisson-Boltzmann 
equation using the lattice Boltzmann method}
For our numerical calculations we developed a lattice Boltzmann method
who's stationary state solves the Poisson-Boltzmann equation in the 
unit cell of the lamellar lipoplex geometry. The lattice Boltzmann method 
simulates densities $f_i$ on a lattice. Each of these densities is 
associated with a velocity ${\bf v}_i$ and they evolve with (scaled) time $t$ 
through the lattice Boltzmann equation 
\begin{equation}
f_i({\bf r}+{\bf v}_i,t+1)=f_i({\bf r},t)+\frac{1}{\tau({\bf r})} [f_i^0({\bf r},t)-f_i({\bf r},t)].
\label{LB}
\end{equation}
We identify the potential $\Psi({\bf r},t)=\sum_i f_i({\bf r},t)$ and
choose the distribution $f_i^0$ such that it has the moments
\begin{eqnarray}
\sum_i f_i^0&=&\tau({\bf r}) \frac{\theta({\bf r}) \kappa^2\sinh(\Psi)+2\pi l_B \rho_f({\bf r})}{6}-\Psi({\bf r},t),
\nonumber \\
\sum_i f_i^0 {\bf v}_i&=& 0,\\
\sum_i f_i^0 {\bf v}_i {\bf v}_i&=& \frac{\Psi(x,t)}{3} \mathbf{1}.\nonumber 
\end{eqnarray} 
We further choose $\tau({\bf r})=[\epsilon({\bf r})/\epsilon_W+1]/2$. 
Summing equation (\ref{LB}) we obtain the macroscopic equation
\begin{equation} \label{gg6}
\frac{\partial \Psi({\bf r},t)}{\partial t} +\nabla \left(\frac{
\epsilon({\bf r})}{\epsilon_W} \nabla \Psi({\bf r},t)\right) = \theta({\bf r}) \kappa^2
\sinh\left(\Psi({\bf r},t)\right)+\rho_f({\bf r})
\end{equation}
so that the stationary solutions of this simulation solves the
Poisson-Boltzmann equation for fixed input charge density 
$\rho_f({\bf r})$ and a given field of dielectric constants 
$\epsilon({\bf r})$. The step function $\theta({\bf r})$ determines the
accessibility of the space to salt ions. Further details on the method 
will be given elsewhere.

\begin{figure}[ht] 
\begin{center}
\includegraphics[width=7cm]{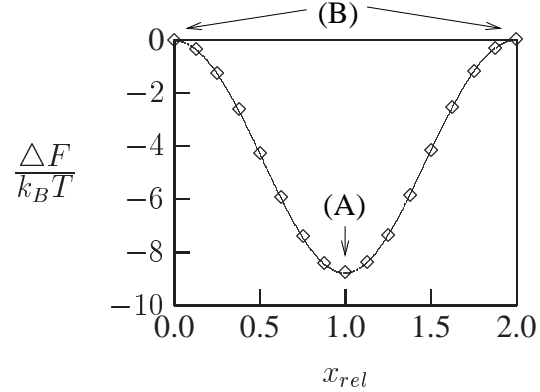}
\end{center}
\caption{\label{FigFree} Numerical results (derived for the same parameters 
as those in Fig.~\ref{FigSIM}) for the change of the free 
energy $\triangle F$ with the relative 
displacement $x_{rel}$ of the DNA arrays with respect to each other. 
The free energies corresponding to the systems shown in 
Fig.~\ref{FigSIM} are indicated by (A) and (B). 
} 
\end{figure}

We set up a simulation with two lipid bilayers and two DNA molecules
imposing periodic boundary conditions. In our simulations one lattice
spacing corresponds to $1\:\mbox{\AA}$. We exclude salt ions from the
membrane regions and from the interior of the DNA (this is achieved by 
the step function $\theta({\bf r})$ in Eq.~\ref{gg6}) and calculate the 
free energy $\triangle F(x_{rel})$ as a function of the scaled lateral 
displacement $x_{rel}$ of the DNA arrays in consecutive galleries of the 
lipoplex. A snapshot of the potential $\Psi({\bf r})$ calculated by 
our simulations is shown in Fig.~\ref{FigSIM}.
The resulting free energy per unit length is then
multiplied with the persistence length of the DNA, $\xi=500\:\mbox{\AA}$, to 
give the change in free energy with the displacement $x_{rel}$ in units of 
$k_BT$. This result is shown in Fig.~\ref{FigFree} and is in good 
agreement with the theoretical prediction of Eq.~\ref{ko6}. More specifically,
Fig.~\ref{FigFree} predicts a maximal change of $\triangle F \approx -9 k_BT$
which happens to coincide with the corresponding value of curve (b) 
in Fig.~\ref{xa2} for $l_D=100 \mbox{\AA}$. 

To sum up, the validity of our approximate analytical model for the 
electrostatic coupling of the two monolayers in a lipid membrane is 
supported by the numerical solutions of the Poisson-Boltzmann 
equation (based on the lattice Boltzmann method). Generally, the
electrostatic communication across a lipid bilayer is quite weak due
to the large drop of the dielectric constant inside the membrane. 
However, as we have shown for a lamellar lipoplex, it cannot always 
be neglected.

\section*{Conclusions}
The electrostatic interaction of a macroion
with an oppositely charged lipid bilayer depends on the coupling parameter 
$H=\epsilon_L l_D/(\epsilon_W d)$. In the (hypothetical) limit $H=0$ 
the interaction reduces to that between the macroion and a single 
monolayer. Here, the apposed monolayer -- separated 
from the macroion by a slab of low dielectric constant -- can be ignored
completely. Differently expressed, for $H=0$ the two monolayers of a 
lipid bilayer are electrostatically decoupled. 

For non-vanishing values of $H$ the behavior of the macroion-membrane complex
is influenced by the electrostatic interaction through the hydrocarbon core 
of the lipid membrane. We have studied this aspect on the basis of 
Gouy-Chapman theory, solving a one-dimensional Poison-Boltzmann equation.
(Thus, all approximations inherent in the Poison-Boltzmann 
formalism are also present in this work). We find interesting and non-trivial 
behavior for non-vanishing $H$, characterized by three different regimes: 
A single weakly charged macroion does not affect the surface potentials on
either one of the two membrane leaflets. As the surface charge of the macroion 
is increased, a regime of close contact will be passed, leading ultimately
to a third regime of high macroion charge where the sign of the 
electrostatic membrane potential, either on one or on both membrane leaflets, 
is reversed. Within our one-dimensional 
model, all characteristic quantities such as membrane potentials,
macroion-membrane distance, free energies, and boundaries between the different
regimes can be calculated analytically.

Typically, the ratio between the
dielectric constants inside the membrane ($\epsilon_L$) and in the water
phase ($\epsilon_W$) is very small, $\epsilon_L/\epsilon_W \approx 1/40$. 
Thus, only for large ratio between the Debye length ($l_D$) and 
the membrane thickness $d$ is $H$ not small compared to $1$. 
Yet, even for $H \ll 1$ electrostatic interactions across a lipid membrane 
can be relevant. To this end, we have estimated the electrostatic 
interaction between the ordered DNA arrays in different galleries of 
the lamellar $L_\alpha^C$ lipoplex structure. For sufficiently large Debye
length $l_D$ we find a significant electrostatic contribution to the
experimentally observed inter-locking of orientationally ordered 
DNA arrays across different galleries. 

Finally, the present work employs the Lattice-Boltzmann method to solve 
the Poisson-Boltzmann equation for the lamellar $L_\alpha^C$ lipoplex 
structure. Our numerical calculations account for the structural details 
of the 
$L_\alpha^C$ lipoplex and compare well with the analytical results 
derived on the basis of the one-dimensional Poisson-Boltzmann model. 
We thus believe that the one-dimensional model is a useful tool to 
estimate interactions between macroions across a lipid bilayer.

\vspace*{0.5cm}

{\bf {\parindent0cm Acknowledgments}} SM would like to thank 
Anne Hinderliter and, during the early stages of this work,
Cristina Baciu for discussions. AJW acknowledges discussions with Eric Foard.
This work was supported by NIH Grant 1 R15 GM077184-01.

\bibliography{BIB}
\bibliographystyle{authordate1}

\end{document}